\documentclass[
aps,nofootinbib,
showpacs,showkeys,preprint
tightenlines,preprintnumbers,] {revtex4}

\usepackage{epsf,epsfig,subfigure,graphicx,amsmath,amssymb 
}
\usepackage{color}
\newcommand{\dis}[1]{\begin{equation}\begin{split}#1\end{split}\end{equation}}
\newcommand{\ie}{{\it i.e.~}}

\newcommand{\gev}{\,\textrm{GeV}}
\newcommand{\meV}{\,\mathrm{MeV}}

\newcommand{\eV}{\,\mathrm{eV}}
\newcommand{\Mp}{M_{\rm P}}

\newcommand{\alem}{\alpha_{\rm em}}

\newcommand{\UDE}{U(1)$_{\rm DE}$}  
  
\newcommand{\UPQ}{U(1)$_{\rm PQ}$}

\def\sw0{{$\sin^2\theta_W^0$}}
\def\thetab{$\bar\theta$}

\def\Nf2{{\bf N_{[2]}}}

\newcommand{\Z}{{\bf Z}}

\def\SU2Ch{SU(2)$_L\times$SU(2)$_R$}
\def\E6{{\rm E_6}}

\def\EE8{{\rm E_8\times E_8'}}

\def\one{{\bf 1}}

\def\one{\bf 1}

\begin{document}

\draft

\title{\Large\bf Quintessential Axions from a New Confining Force
}

\author{Jihn E.  Kim, Younggeun Kim and Soonkeon Nam}
\address
{Department of Physics, Kyung Hee University, 26 Gyungheedaero, Dongdaemun-Gu, Seoul 02447, Republic of Korea  
}
 
\begin{abstract} 
Almost massless mesons created at the condensation scale of extra nonabelian gauge group can be  candidates of dark energy source. This can be another motivation for introducing an additional confining force.  

\end{abstract}
\maketitle

\section{Introduction}\label{sec:Introduction}

The origin of dark energy (DE) in the Universe is most mysterious in the current cosmology. At present, the DE density $(0.003\,\eV)^4$ is synonymous to the cosmological constant (CC). In the multiverse distinguished by CC, the anthropic choice allows some universes to evolve to the ones similar to our currently nearly flat and empty Universe to have creatures to quest about it now \cite{Weinberg88}. On the other hand, if the current DE is a transient one, parametrized by a varying $\theta$,
the classically evolving scalar field  $\theta$ is interpreted as the source for the current  domination of the energy density of the Universe
\cite{FriemanWaga98,Carroll98}. The name `quintessential axion' \cite{KimNilles03,Hill02,KimRMP10} is by identifying $\theta$ as the pseudo-Goldstone boson of spontaneously broken global symmetry \UDE.

Light pseudoscalar particles are the lampposts to the road leading to physics scales much above their masses. The well-known example is the pion triplet looked at low energy strong interaction scale, which guided toward the \SU2Ch~chiral symmetry at a higher energy scale \cite{Nambu61,GellMann68}. Another is the very light axion which hints at the intermediate scale \cite{KimPRL79,KNS18}. In this
 pseudoscalar scenario, the scale where the symmetry is explicit is at the scale
$E\gtrsim f\sim\Lambda^2/m$, for the pseudoscalar mass $m$, where $\Lambda^4$ is a typical energy density contributed by such a pseudoscalar. Before obtaining mass of quintessential axion by the energy density  perturbation of order $\Lambda^4$, the pseudoscalar degree is the phase degree $\theta(x)$ of some unitary operators such that they have kinetic energy terms in quantum field theory below the defining scale $f$ of the pseudoscalar $a$. Below the scale $f$, the global transformation is expressed as  $\theta\to a/f$ where   $f$  is   the {\it decay constant}. Above the scale $f$, the phase in $e^{i\,a/f}$  is not a dynamical field, \ie not depending on $x$, but it still represents a phase direction of the global symmetry.
On the other hand, if a pseudoscalar is originally present in the theory, then $f$ is the defining scale of that theory. In this sense, string scale $M_s$ itself is the decay constant for axions from string theory.  

Quintessential axion requires \cite{Carroll98}:
\begin{itemize}
\item The decay constant is near the Planck scale $\Mp$, and
\item Mass is about $10^{-32\,}\eV$.
\end{itemize}
Since the Planck mass is the mass defining scale, the first condition looks easy to be implemented. But, the  singlets beyond the standard model  may be required to interact with light fields, in which case a judicious care is needed to allow the VEV of singlets still remaining near the Planck scale. Implementing the second condition needs some symmetries such that the breaking scale is sufficiently small.
 
If the pseudoscalar mass $m$ is less than 1 MeV, it can rarely decay to two photons and two neutrinos. Using the $\pi^0$ decay rate to two photons   $\Gamma(\pi^0\to 2\gamma)\simeq ( \alem^2/64\pi^3) (M_{\pi^0}/f_\pi)^2M_{\pi^0}$ (at leading order \cite{Holstein02}) and   to two neutrinos  $1.2\times 10^{-5}\Gamma(\pi^0\to 2\gamma)$ \cite{Shrock79}, we can get an idea on the lifetime of the pseudoscalar.  For it to be still present in our Universe, therefore we require $\Gamma^{-1} > 4.3\times 10^{17}{\rm s} = 1/(1.53\times 10^{-33} \eV)$.
 Using the above $f-m$ relation, we get an idea on $m$ from the condition that it survives until now, using the  QCD scale $\Lambda$,
 \dis{
 \Gamma^{-1}\simeq  \frac{ 64\pi^3\,\Lambda_{\rm QCD}^4}{\alem^2m^4}   \frac{1}{m}>\frac{1}{1.53\times 10^{-33} \eV}
 }
or $m< 65\, \eV$ for $\Lambda_{\rm QCD}=380\,\meV$.\footnote{For the very light axion, the specific coupling reduces it further to 24\,eV \cite{KimRMP10}.} If we take $\Lambda^4$ as the current energy density of the Universe,  then pseudoscalars with mass less than $3.3\times 10^{-4}\, \eV$ survives until now.\footnote{Ultra-light axions (ULAs) \cite{KimPRD16,Witten17} was suggested for the galactic scale structures.}  
 
 We will be using two words for describing breaking the global symmetry. One is the spontaneous breaking scale $f$ mentioned above. This is the vacuum expectation value (VEV) of the scalar field which  carries the non-vanishing quantum number of the global symmetry and is determined from the interaction terms respecting the global symmetry. The other is the {\it explicit} breaking terms which are interaction terms explicitly breaking the global symmetry in question and provide the curvature in the direction of $a$.

\color{black}

 We will consider the QCD axion also because it is present in most models.  Not only providing a solution of the strong CP problem,  the QCD axion can work as a source of dark matter (DM) in the Universe \cite{PWW83,AS83,DF83,Bae08},
  which depends on the confining scale  $\Lambda_{\rm QCD}$ and the axion decay constant $f_a$.
So, let us start with building a field theoretic model housing both the QCD  and quintessential axions. Working out an explicit example, we employ  supersymmetry because supersymmetry reduces the number of couplings and hence the discussion for  generating symmetry  breaking terms is much simpler.

Let the gauge group for a new confining force be $\cal G$ and  the flavor symmetry of ex-quarks under  $\cal G$ be U$(N)\times$U$(N)$ which reduces to SU$(N)\times$SU$(N)$ by removing the heavy singlet meson obtaining mass by the anomaly contribution.  The light mesons $\Pi^i_j(i=1,\cdots,N;j=1,\cdots,N)$ with Tr\,$\Pi=0$, belonging to the adjoint representation ($\bf N^2-1$) of SU$(N)_A\subset$ SU$(N)\times$SU$(N)$, obtain mass by explicit breaking of the flavor symmetry of  ex-quarks.

One obvious scale breaking the global symmetry is the condensation scale $\Lambda$ of the ex-quarks,
 \dis{
\langle \overline{Q}_LT^{a\,i}_jQ_L \rangle=\Lambda^3 e^{i\Pi^{ai}_j/f}
 }
where $\overline{Q}_LT^{a\,i}_jQ_L $ is interpreted as the matrix for adjoint representation $T^a$ sandwiched between the relevant fermion fields $\overline\psi_R T^a\psi_L$. To construct a model based on discrete symmetry, supersymmetric extension is simpler and we will discuss supersymmetric extension below.
  
In the ${\cal N}=1$ supersymmetric extension, we include scalar partners of ex-quarks also.  These scalar quarks are defined to have  the chirality of the corresponding ex-quarks denoted by tilded fields.  The SU$(N)_L\times$SU$(N)_R$ representation of scalar ex-quarks are
\dis{
Q_L=\begin{pmatrix} a_1\\ a_2\\  \cdot\\ \cdot   \\ \cdot\\  a_N\end{pmatrix}_L,~~~~\overline{Q}_L=\begin{pmatrix} \bar{b}_1\\ \bar{b}_2\\  \cdot\\ \cdot   \\ \cdot\\  \bar{b}_N\end{pmatrix}_L.
}
Usually the same notations $Q_L$ and $Q_R$  are used for the corresponding superfields also, $Q_L+\sqrt2\tilde{Q}_L\,\vartheta+F_{Q_L} \vartheta^2$, etc., and interaction is given by $\int d^4xd^2\vartheta W$.
In supersymmetric extension, the  R-charges of chiral superfields determine possible terms in the superpotential $W$. For dynamical SUSY breaking via the new confining gauge group ${\cal G}$, the gaugino condensation scale $\Lambda$ is expected around $10^{13\,}\gev$ \cite{Nilles82}. With \UDE, we also consider  the breaking scale of \UPQ~for the QCD axion \cite{KNS18}. For simplicity, let us assume the VEV of the singlet scalar field $\langle\sigma\rangle(\equiv V)$ is the QCD axion scale $f_a\approx 10^{10\,}\gev$.  In addition to singlet fields, below we will consider Higgs fields also. 

In supersymmetric models, condensation is possible for bilinears of fermions and also bilinears of scalar ex-quarks (bi-scalar). Condensation of  bi-scalars does not break supersymmetry.   This is a merit introducing quintessential axion in the phases of bi-scalar condensation.  So, $f$ for the  bi-scalar condensation can be different from the gaugino condensation scale $\Lambda$.

 Condensation of bi-scalars is expected when the gauge coupling of the new non-abelian force becomes strong at $\Lambda$. Let the bi-scalar singlet be $X$,
\dis{
\overline{Q}_LQ_L\equiv X.
}
As mentioned above, nonzero $X$ does not break supersymmetry. If we consider a superpotential in terms of $X$, 
\dis{
&W= \Lambda X -\frac{1}{2M} X^2+\cdots.\\
}
where $ \Lambda$ and $1/M$ can result from the VEVs of some singlet fields and $M$ in the following is used for the Planck mass $\Mp=2.43\times 10^{18\,}\gev$. So,  
the first minimum of the potential appears from $V= \left(\Lambda- \frac{1}{\Mp}X\right)^2 $ and the next minimum can be far above the Planck scale.
Then, the decay constant $f=\sqrt{ X}$ is expected at a median of $\Lambda$ and $\Mp$.
 
Now let us consider the explicit breaking terms. Anomalies of non-abelian gauge groups are inevitable breaking terms, and the QCD anomaly is used for the explicit breaking term of the \UPQ~global symmetry. If we consider a phase of the {\it SM singlet scalar field}, the weak gauge group in the SM may work for an explicit breaking term of the \UDE~global symmetry \cite{Kim21PLB}. But our quintessence fields are mesons in the phase of bi-scalars, or scalar ex-quarks, and hence the SU(2)$_W$ anomaly cannot work for our quintessence fields. The only method is from  the superpotential made of scalar fields. To have a sufficiently small numerical value for the vacuum energy, we consider the VEV scale of the SM Higgs fields. A possible superpotential is
\dis{
\Delta W=\frac{1}{M^{n+3}}\overline{Q}_LQ_L\left( H_u H_d\right)^2\sigma^{n},\label{eq:DelW6}
}
where we started from $\left( H_u H_d\right)^2$.
If we started with one power of $H_uH_d$, too large a value of $n$ would be required.  
 Considering condensation of the hidden sector quark $Q$ in Eq. (\ref{eq:DelW6}), the vacuum energy density  is
\dis{
 \frac{4}{M^{n+3}}\Lambda^3 (v_uv_d)^2V^n= \frac{4}{M^{n+3}}\Lambda^3 \frac{v_{d}^4}{\cos\beta^4}V^n\simeq (0.003\eV)^4\label{eq:DelV6}
}
where $\tan\beta=v_u/v_d$. 

\begin{figure}[!t]\hskip -0.3cm
\includegraphics[height=0.45\textwidth]{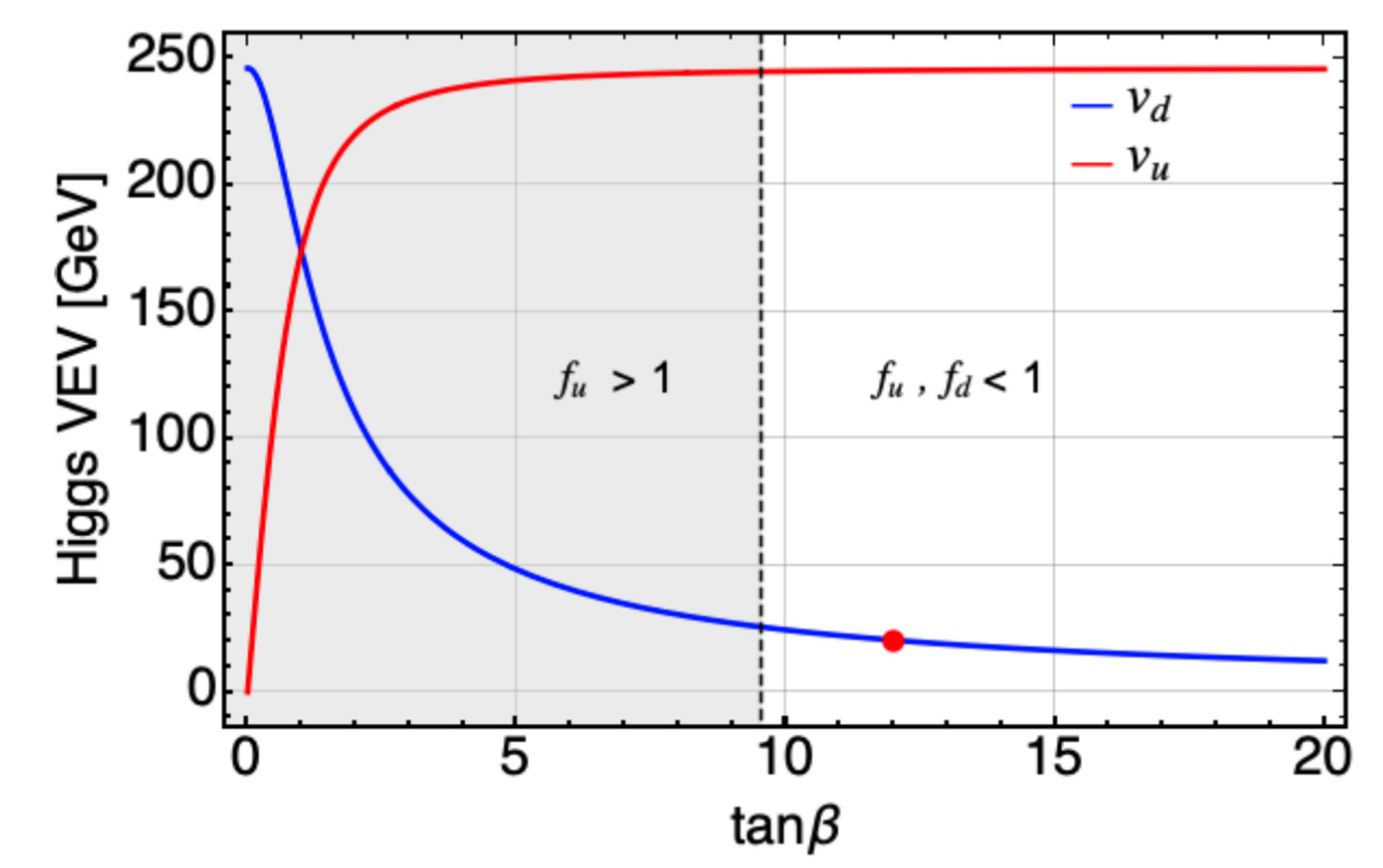}
\caption{ Potential generated by Yukawa terms breaking \UDE. At the intersection of the blue curve and the $f_u=1$ line, $v_d$ is $25.6\gev$.}\label{fig:YukawaLoop}
\end{figure}
The top quark mass $m_t$ is $f_u v_u/\sqrt2$  and the bottom quark mass $m_b$ is $f_d v_d/\sqrt2$.  For the perturbative calculation to be possible, we limit the study in the region $f_u\le 1$ and $f_d\le 1$. Equation (\ref{eq:DelV6}) has one parameter $\tan\beta$.  For a given  $\tan\beta$, relative values of $f_u$ and $f_d$ are given, as shown in Fig. \ref{fig:YukawaLoop}.   The blue curve in Fig. \ref{fig:YukawaLoop} is interpreted as the current DE, $(0.003\eV)^4$, for $f_d=1$.

 As an example, taking $v_d=20\gev$ and  $m_t=173\,\gev$,  Eq. (\ref{eq:DelV6}) is estimated as,
\dis{ 
 \frac{4}{M^{n+3}}\Lambda^3 \frac{v_{d}^4}{\cos^4\beta}V^n\simeq \tan^4\beta\left(\frac{V}{M}\right)^n \left(\frac{20}{2.43\times 10^{18}} \right)^4 \Lambda^3 \cdot 0.97\times 10^{28}\eV=10^{-10}\eV^4,
 }
 leading to
\dis{
\Lambda =  \left\{\begin{array}{l}
2.7\times 10^{+14}\gev    , ~\textrm{for }n=1\\ 1.7\times 10^{+17}\gev    , ~\textrm{for }n=2\\
\end{array}
\right.\label{eq:ns}
}
This example is shown as the red bullet in   Fig. \ref{fig:YukawaLoop}.

\begin{table}[t!]
\begin{center}
\begin{tabular}{@{}l|c|c|cc@{}} \toprule
 & Representation under ${\cal G}\equiv$SU(${\cal N}$)  & SU$(2)_W\times$U(1)$_Y$ &${\Z}_{6R}$ 
  \\[0.2em]\hline
$Q_L$  & ${\cal N}$& ${\one}$ &$+1$ &    \\[0.2em]
$\overline{Q}_L$  & $\overline{\cal N}$&  ${\one}$ & $-1$ &    \\[0.2em]
$H_u$  &${\one}$& ${\bf 2}_{+1/2}$ &$+3$  &   \\[0.2em]
$H_d$  & ${\one}$ &  ${\bf 2}_{-1/2}$ &$+2$  &   \\[0.2em]
$\sigma$  & ${\one}$& ${\one}$ &$+4$ &    \\[0.2em]
$S$  & ${\one}$& ${\one}$ &$+5$ &    \\[0.2em]
\botrule
\end{tabular} 
\end{center}
\caption{${\Z}_{6R}$ quantum numbers of relevant chiral superfileds appearing in Eq. (\ref{eq:DelW6}).}\label{tab:SUSY6} 
\end{table}

In Table \ref{tab:SUSY6}, an example for the ${\Z}_{6R}$ discrete symmetry is shown, from which
 $n=1$ is chosen in Eq. (\ref{eq:ns}).  The ${\Z}_{6R}$ is a discrete subgroup of R-symmetry U(1)$_R$.  Because the antisymmetric parameter $\vartheta$ carries --1 unit of   U(1)$_R$ charge, superpotential $W$ with +2 units of  U(1)$_R$ charge survives in the integration $\int d^2\vartheta\,W$.  To assign tree level Yukawa couplings of the SM quarks, we have the following
quantum numbers  
 \dis{
 q_L: 0, ~ u_L^c: -1,~ d_L^c:0,
 }
 such that both $q_Lu_L^c H_u$ and  $q_Ld_L^c H_d$ carry two units (modulo 6) of U(1)$_R$ charge.
 To obtain a TeV scale $\mu$ term \cite{KimNilles84}, one can introduce the following superpotential 
\dis{
W_\mu=\frac{(10^{10\,}\gev)^2}{M}\,H_uH_d,\label{eq:muneeded}
} 
but should forbid the dimensions 2 and 3 superpotential terms $H_uH_d,H_uH_d\sigma$ and $H_uH_dS$. Equation (\ref{eq:muneeded}) can be satisfied with $\langle S\rangle\approx \langle \sigma\rangle\approx 10^{10\,}\gev$.  Then, singlets $S$ and $\sigma$ lead to the following superpotential
\dis{
W= -\alpha\sigma S^2 +\frac{\varepsilon}{M} S^4-\frac{x}{M^2}\sigma S^2 Q_L\overline{Q}_L+\cdots ,\label{eq:WSss}
}
where the neglected terms $\cdots$ denote other terms including $Q_L\overline{Q}_L$. 
The relations between  the VEVs of $\sigma$ and $S$ are
\dis{
&\frac{\partial W}{\partial \sigma}:\to Q_L\overline{Q}_L=-\frac{\alpha M^2}{x}\\ 
&\frac{\partial W}{\partial S}:\to (x\frac{Q_L\overline{Q}_L}{M^2}+\alpha )\sigma=\frac{2\varepsilon  }{M} S^2 .\label{eq:SQQ}
}
Relations in (\ref{eq:SQQ}) give exact SUSY. But SUSY is broken dynamically by the condensation of ex-quarks at $\Lambda \simeq 10^{13\,}\gev$. This SUSY breaking effect is added in the first SUSY relation,
\dis{
-\alpha  S^2 -\frac{x}{M^2}  S^2 Q_L\overline{Q}_L +\delta_1 \Lambda^2 =0,\\
\frac{2\varepsilon S^3}{M} -\alpha  S\sigma -\frac{x}{M^2}  S Q_L\overline{Q}_L\sigma+\frac{\delta_2\Lambda^2}{2}=0 ,\\
2\frac{\varepsilon}{M} S^4+(\frac{\delta_2S-2\delta_1\sigma}{2})\Lambda^2=0.\label{eq:SUSYrel}
 }
Solutions of $\sigma$ and $S$ satisfying Eq. (\ref{eq:SUSYrel}) are shown in Fig. \ref{fig:YukawaLoop}.
So, the needed $\mu$ term is
around the point $(1,1)$ in Fig. \ref{fig:YukawaLoop}.

The above U(1)$_R$ symmetry is a kind of the PQ symmetry because the  U(1)$_R$ quantum number of $H_uH_d$ is non-zero. At the scale $V_{\rm PQ}\equiv\sqrt{\sigma^2+S^2}$, therefore, the PQ symmetry is spontaneously broken and there results the QCD axion discussed in the literature \cite{KimRMP10}. Because the quantum numbers of $\sigma$ and $S$ are relatively prime, all the axionic vacua are connected and the domain wall number is 1 \cite{KNS18,ChoiKim85}. This QCD axion is responsible for the DM in the Universe.  On the other hand, the mesons resulting from the new confining force are responsible for the DE of the Universe.

\begin{figure}[!t]\hskip -0.3cm
\includegraphics[height=0.55\textwidth]{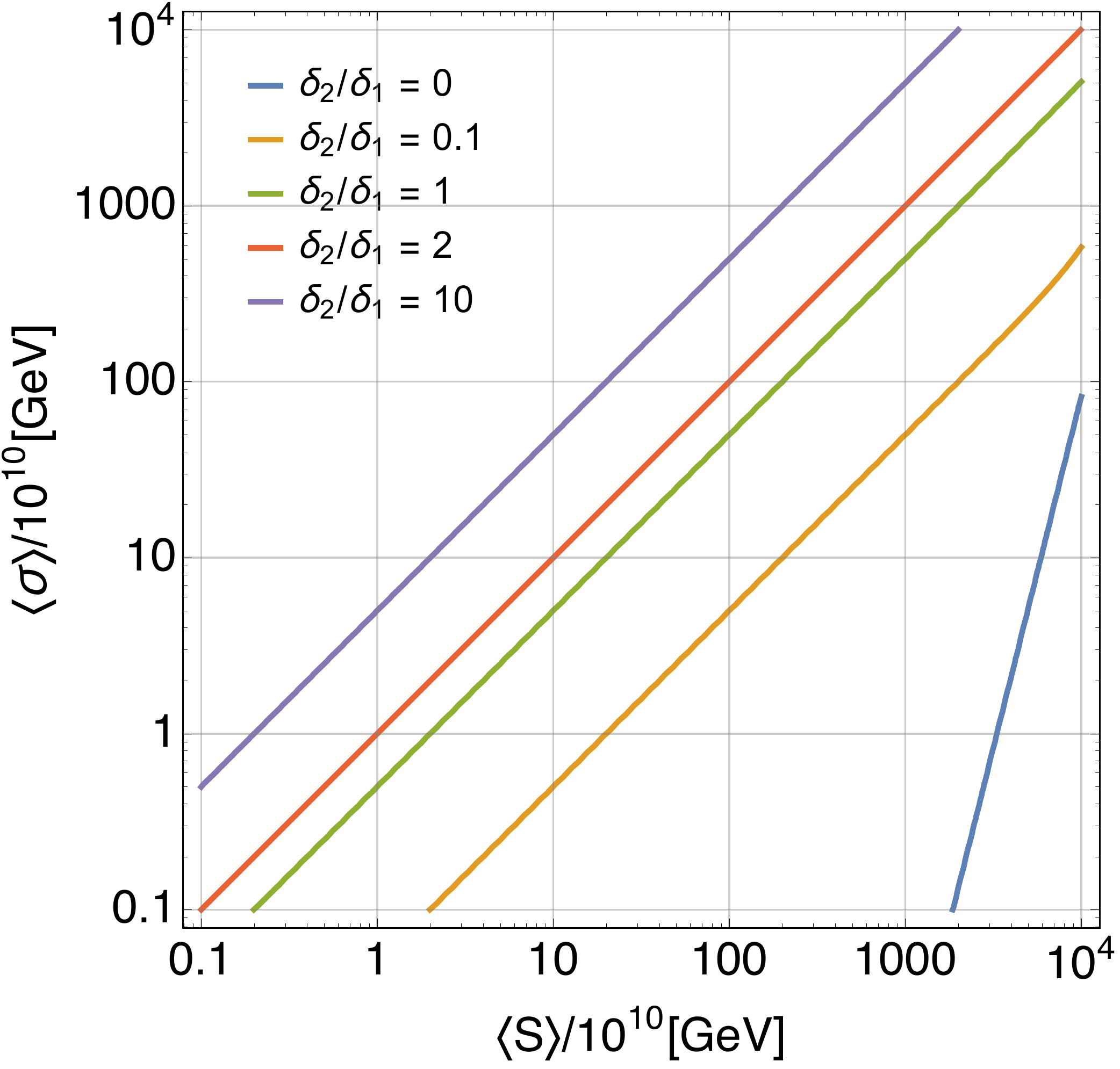}
\caption{Solutions of $\sigma$ and $S$ satisfying Eq. (\ref{eq:SUSYrel}).}\label{fig:YukawaLoop}
\end{figure}

\begin{figure}[!t]\hskip -0.3cm
\includegraphics[height=0.13\textwidth]{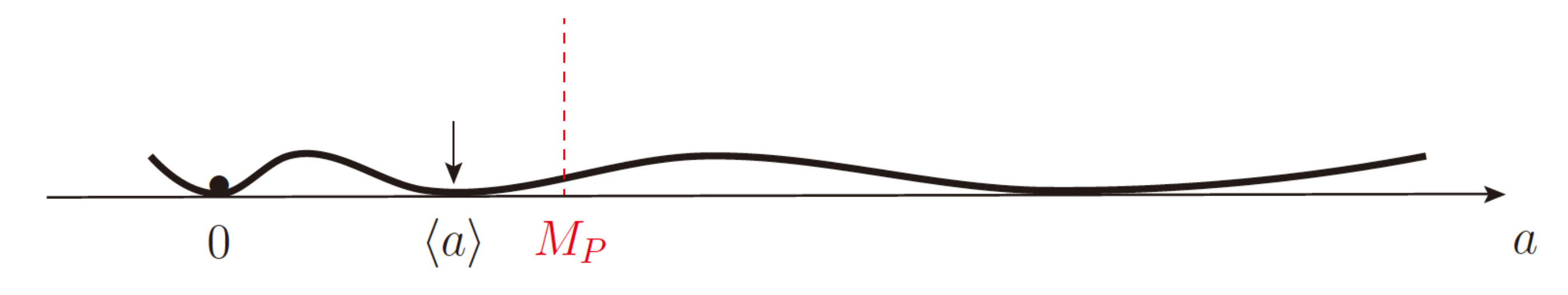}
\caption{Shape of the potential by Eq. (\ref{eq:Vterms}).}\label{fig:PotMp}
\end{figure}

Gravity effects of breaking global and discrete symmetries can be given by the terms explicitly breaking them. Since supergravity with superpotential $W$ has a fine-tuning problem of adjusting the cosmological constant \cite{Nilles04}, we lack any understanding on the magnitude of the cosmological constant in supergravity.  So, no attempt is made to introduce  explicit DE terms. However, we can discuss the gravity effects to the parameters in the gauge part, especially on the \thetab~of QCD, as done in \cite{Barr92}. Here, we consider only the potential $V$  satisfying the $\Z_{6R}$ symmetry, but not the superpotential $W$. Possible  $\Z_{6R}$ invariant terms are
\dis{
\sigma^3,~{\sigma^*}^3,~S^2{\sigma^*} ,~S^6,~{S^*}^6,~(\sigma\sigma^*)^n~{\rm for}~n=1,2,\cdots,~(SS^*)^n~{\rm for}~n=1,2,\cdots, ~{\rm etc.}\label{eq:Vterms}
}
Note that the $(\sigma\sigma^*)^n$ and $(SS^*)^n$ terms respect the PQ symmetry and do  not   shift  \thetab. The leading terms in shifting \thetab~in Eq. (\ref{eq:Vterms}) are the cubic terms. But these cubic terms do not arise from the superpotential and hence appear with supersymmetry breaking coefficients, \ie the gaugino condensation $\langle {\cal G} {\cal G}\rangle$ of the new confining force. Thus the \thetab~shift is estimated from
\dis{
\frac{1}{M^2}\langle {\cal G} {\cal G}\rangle\sigma^3,\cdots
}
which can be compared to the axion (or Pontryagin number) shift by $\langle G_{\mu\nu}\tilde{G}^{\mu\nu} \rangle\approx \Lambda^4$,
\dis{
\Delta\bar{\theta}\approx  \frac{\sigma^3}{M^2\Lambda}\approx 1.7\times 10^{-20},
}
for $\Lambda=10^{13}\gev$ and $\sigma=10^{10}\gev$. So, the \thetab~shift by gravity effects falls in the allowed region, $|\bar{\theta}| \lesssim 10^{-11}$  \cite{ KimRMP10}. In addition, the higher order terms will give a potential of the form given in Fig. \ref{fig:PotMp}. Restrcting the vaccum for $a$ below $\Mp$ as shown in the left-hand side of the dash line, the nonzero VEV of $a$ is near the Planck scale. In addition, note that the effects of pseudo-Goldstone bosons may induce terms expected in general relativity\cite{Bludman77}. We note, however, that all such effects with supersymmetry  are included in Eq. (\ref{eq:WSss}). 

In conclusion, we propsed a mechanism creating quintessential axions that are almost 
massless mesons created at the condensation scale of a new nonabelian gauge group.  Since an extra nonabelian gauge group has been proposed for dymnamical supersymmetry breaking, our mechanism is better suited in supersymmetric models.

\color{black}
 
\acknowledgments{This work is supported in part by the National Research Foundation (NRF) grant  NRF-2018R1A2A3074631. } 



\begin{thebibliography}{99}
\def\prp#1#2#3{{Phys.\,Rep.}  {\bf #1} (#3) #2}
\def\rmp#1#2#3{{ Rev. Mod. Phys.}  {\bf #1} (#3) #2}
\def\npb#1#2#3{{ Nucl.\,Phys.\,B}   {\bf #1} (#3) #2}
\def\plb#1#2#3{{Phys.\,Lett.\,B}   {\bf #1} (#3) #2}
\def\prv#1#2#3{{Phys.\,Rev.}  {\bf #1} (#3) #2}
\def\prd#1#2#3{{Phys.\,Rev.\,D}  {\bf #1} (#3) #2}
\def\prl#1#2#3{{ Phys.\,Rev.\,Lett.}   {\bf #1} (#3) #2}
\def\err#1#2#3{ {\bf #1}   {\bf #1} (#3) #2\,(E)}
\def\jhep#1#2#3{{ JHEP}   {\bf #1} (#3) #2}
\def\jcap#1#2#3{{ JCAP}   {\bf #1} (#3) #2}
\def\zp#1#2#3{{ Z.\,Phys.}  {\bf #1} (#3) #2}
\def\epjc#1#2#3{{ Euro.\,Phys.\,J.\,C}  {\bf #1} (#3) #2}
\def\jpg#1#2#3{{J.\,Phys.\,G}   {\bf #1} (#3) #2}
\def\ijmpa#1#2#3{{ Int.\,J.\,Mod.\,Phys.\,A}   {\bf #1} (#3) #2}
\def\mpla#1#2#3{{Mod.\,Phys.\,Lett.\,A}   {\bf #1} (#3) #2}
\def\apj#1#2#3{{Astrophys.\,J.}   {\bf #1} (#3) #2}
\def\nat#1#2#3{{Nature}   {\bf #1} (#3) #2}
\def\sjnp#1#2#3{{ Sov.\,J.\,Nucl.\,Phys.}  {\bf #1} (#3) #2}
\def\apj#1#2#3{{Astrophys.\,J.}   {\bf #1} (#3) #2}
\def\mnra#1#2#3{{ Mon.\,Not.\,Roy.\,Astron.\,Soc.}   {\bf #1} (#3) #2}
\def\jetpl#1#2#3{{JETP\,Lett.}   {\bf #1} (#3) #2}
\def\pthp#1#2#3{{Prog.\,Theor.\,Phys.}  {\bf #1} (#3) #2}
\def\jkps#1#2#3{{J.\,Korean\,Phys.\,Soc.}  {\bf #1} (#3) #2}
\def\dum#1#2#3{  {\bf #1} (#3) #2}

\def\ibid#1#2#3{{\it ibid.} {\bf #1} (#3) #2}
\def\err#1#2#3{\ {\bf #1} (#3) #2\,(E)}

\bibitem{Weinberg88} S. Weinberg, \emph{Anthropic Bound on the Cosmological Constant}, \prl{59}{2607}{1987} [doi: 10.1103/PhysRevLett.59.2607].

\bibitem{FriemanWaga98} J. A. Frieman and I. Waga, \emph{Constraints from high redshift supernovae upon scalar field cosmologies}, \prd{57}{4642}{1998} [doi: 10.1103/PhysRevD.57.4642].

\bibitem{Carroll98} S. M. Carroll, \emph{Quintessence and the rest of the world}, \prl{81}{3067}{1998} [e-Print: astro-ph/9806099 [astro-ph]].

\bibitem{KimNilles03} J. E. Kim and H. P. Nilles, \emph{A quintessential axion}, \plb{553}{1}{2003} [e-print: hep-ph/0210402 [hep-ph]].

\bibitem{Hill02} C. T. Hill, \emph{Natural theories of ultralow mass pseudo Nambu-Goldstone bosons: Axions and quintessence}, \prd{66}{075010}{2002} [e-print:hep-ph/0205237].

\bibitem{KimRMP10} J. E. Kim and G. Carosi, \emph{ Axions and the strong CP problem}, \rmp{82}{557}{2010} [e-print: 0807.3125 [hep-ph]]. 

\bibitem{Nambu61} Y. Nambu and G. Jona-Lasinio, 
\emph{Dynamical Model of Elementary Particles Based on an Analogy with Superconductivity. 1.},  \prd{122}{345}{1961} [doi:10.1103/PhysRev.122.345]. 

\bibitem{GellMann68} M. Gell-Mann, R. J. Oakes, and B. Renner,
\emph{Behavior of current divergences under SU(3)$\times$SU(3)},  \prd{175}{2195}{1968} [doi:10.1103/PhysRev.175.2195]. 

\bibitem{KimPRL79} J. E. Kim, \emph{Weak Interaction Singlet and Strong CP Invariance}, \prl{43}{103}{1979} [doi:10.1103/ PhysRevLett.43.103].

\bibitem{KNS18} For a review, see, J. E. Kim, S. Nam, and Y. Semertzidis, \emph{Fate of global symmetries in the Universe: QCD axion, quintessential axion and trans-Planckian inflaton decay-constant}, \ijmpa{33}{1830002}{2018} [e-print: 1712.08648 [hep-ph]].

\bibitem{Holstein02} J. L. Goity, A. M. Bernstein, and B. R. Holstein, \emph{Decay $\pi^0\to \gamma\gamma$ to next to leading order in chiral perturbation theory}, \prd{66}{076014}{2002} [e-print:hep-ph/0206007].

\bibitem{Shrock79} R. E. Shrock and M. Voloshin, \emph{Bounds on quark mixing angles from the decay $K_L\to \mu\bar{\mu}$}, \plb{87}{375}{1979} [doi:10.1016/0370-2693(79)90557-4].


\bibitem{KimPRD16} J. E. Kim and D. J. E. Marsh, \emph{An ultralight pseudoscalar boson}, \prd{93}{025027}{2016} [e-Print: 1510.01701 [hep-ph]].

\bibitem{Witten17} 
L. Hui, J. P. Ostriker, S. Tremaine, and E. Witten,
\emph{Ultralight scalars as cosmological dark matter}, \prd{95}{043541}{2017} [e-print:1610.08297 [astro-ph.CO]].

\bibitem{PWW83} J. Preskill, M. B. Wise, and F. Wilczek, \emph{Cosmology of the Invisible Axion}, \plb{120}{127}{1983} [doi:10.1016/0370-2693(83)90637-8].

\bibitem{AS83} L. F. Abbott and P. Sikivie, \emph{A Cosmological Bound on the Invisible Axion}, \plb{120}{133}{1983} [doi:10.1016/0370-2693(83)90638-X].

\bibitem{DF83} M. Dine and W. Fischler, \emph{The Not So Harmless Axion}, \plb{120}{137}{1983} [doi:10.1016/0370-2693(83)90639-1].

\bibitem{Bae08} See a fit, K. J. Bae, J-H. Huh, and J. E. Kim, \emph{Update of axion CDM energy density}, \jcap{09}{005}{2008} [e-print: 0806.0497 [hep-ph]].


\bibitem{Nilles82} H. P. Nilles, \emph{Dynamically Broken Supergravity and the Hierarchy Problem}, \plb{115}{193}{1982} [doi: 10.1016/0370-2693(82)90642-6].


\bibitem{Kim21PLB} J. E. Kim, \emph{Anomalies and parities for quintessential and ultra-light axions}, \plb{817}{136248}{2021} [e-Print: 2102.01795 [hep-ph]].




\bibitem{quinessence} M. Bronstein, Phys. Z. Sowjetunion 3 (1933) 73;\\
M. $\ddot{\rm O}$zer, M. O. Taha, \npb{287}{797}{1987} ;\\
B. Ratra and P. J. E. Peebles, \emph{Cosmological consequences of a rolling homogeneous scalar field}, \prd{37}{3406}{1988} [doi:10.1103/PhysRevD.37.3406];\\
C. Wetterich, \emph{Cosmologies with variable Newton's `constant'}, \npb{302}{645}{1988} [doi:10.1016/0550-3213(88)90192-7];\\  
H. Gies and C. Wetterich, \emph{Renormalization flow from UV to IR degrees of freedom}, hep-ph/0205226;\\
J. A. Frieman, C. T. Hill, and R. Watkins, \emph{Late-time cosmological phase transitions: Particle-physics models and cosmic evolution}, \prd{46}{1226}{1992} [doi:10.1103/PhysRevD.46.1226];\\  
R. Caldwell, R. Dave, and P. J. Steinhardt, \emph{Cosmological imprint of an energy component with general equation of state}, \prl{80}{1582}{1998} [doi: 10.1103/PhysRevLett.80.1582];\\   
P. Binetruy,   \emph{Models of dynamical supersymmetry breaking and quintessence}, \prd{60}{063502}{1999} [e-Print: hep-ph/9810553 [hep-ph]];\\ 
C. Kolda and D. H. Lyth, \emph{Quintessential difficulties}, \plb{458}{197}{1999} [e-Print: hep-ph/9811375 [hep-ph]];\\  
T. Chiba, \emph{Quintessence, the gravitational constant, and gravity}, \prd{60}{083508}{1999} [doi: 10.1103/PhysRevD.60.083508];\\
P. Brax and J. Martin,  \plb{468}{40}{1999} [];\\  
A. Masiero, M. Pietroni, F. Rosati,   \emph{SUSY QCD and quintessence}, \prd{61}{023504}{2000} [e-Print: hep-ph/9905346 [hep-ph]];\\
J. E. Kim, 
\emph{Model dependent axion as quintessence with almost massless hidden sector quarks}, \jhep{06}{016}{2000} [e-Print: hep-ph/9907528 [hep-ph]];\\
M. C. Bento, O. Bertolami, Gen. Relativ. Gravit. 31 (1999)
1461 [e-Print: gr-qc/9905075 [gr-qc]];\\
F. Perrotta, C. Baccigalupi, S. Matarrase,  \emph{Extended quintessence}, \prd{61}{023507}{2000} [e-Print: astro-ph/9906066 [astro-ph]];\\
J. E. Kim, 
\emph{Model dependent axion as quintessence with almost massless hidden sector quarks}, \jhep{06}{016}{2000} [e-Print: hep-ph/9907528 [hep-ph]];\\
A. Arbey, J. Lesgourgues, P. Salati, \emph{Cosmological constraints on quintessential halos}, \prd{65}{083514}{2002} [e-Print: astro-ph/0112324 [astro-ph]].

\bibitem{ChoiK21} K. Choi, S. H. Im, and C. S. Shin, \emph{Recent Progress in the Physics
of Axions and Axion-Like
Particles}, Ann. Rev. Nucl. Part. Sci. {\bf 71} (2021) 225 [e-Print: 2012.05029 [hep-ph]].

\bibitem{KNP05JCAP}
 J. E. Kim, H. P. Nilles, and M. Peloso, \emph{Completing natural inflation}, \jcap{01}{005}{2005} [e-Print: hep-ph/0409138 [hep-ph]].


\bibitem{Marsh21} For a recent survay, see, A. Lagu$\ddot{\rm e}$, J. R. Bond, R. Hlozek, K. K. Rogers, D. J. E. Marsh, and D. Grin, \emph{Constraining Ultralight Axions with Galaxy Survaeys}, eprint:2104.07802[astro-ph.CO].

\bibitem{Kibble77} T. W. B.Kibble, \emph{Topology of cosmic domains and strings}, Jour. of Physics A: Mathematical and General {\bf 9} (1976) 1387.  

\bibitem{Sikivie82DW} P. Sikivie, \emph{Of Axions, Domain Walls and the Early Universe}, \prl{48}{1156}{1982} [doi:10.1103/ PhysRevLett.48.1156].

\bibitem{Barr87}  S. M. Barr,  K. Choi and J. E. Kim, \emph{Axion Cosmology in Superstring Models}, \npb{283}{591}{1987} [doi:10.1016/0550-3213(87)90288-4].

 \bibitem{KimNilles84} J. E. Kim and H. P. Nilles, \emph{The $\mu$ problem and the strong CP problem}, \plb{138}{150}{1984} [doi:10.1016/0370-2693(84)91890-2]. 
 
\bibitem{ChoiKim85} K. Choi and J. E. Kim, \emph{Domain Walls in Superstring Models}, \prl{55}{2637}{1985} [doi:10.1103/PhysRevLett.55.2637].

\bibitem{Nilles04} H
 P. Nilles, \emph{Supersymmetry, Supergravity and Particle Physics}, \prp{110}{1}{1984} [doi:10.1016/0370-1573(84)90008-5].
 
 \bibitem{Barr92} S. M. Barr and D. Seckel, \emph{Planck scale corrections to axion models}, \prd{46}{539}{1992} [doi: 10.1103/PhysRevD.46.539].
 
\bibitem{Bludman77} S. A. Bludman and M. A. Ruderman,  \emph{Induced Cosmological Constant Expected above the Phase Transition Restoring the Broken Symmetry}, \prl{38}{255}{1977} [doi:10.1103/PhysRevLett.38.255].
  
\end{thebibliography}
\end{document}